\documentclass{aastex631}

\usepackage{graphicx}	
\usepackage{amsmath}	
\usepackage{makecell}

\newcommand\wise{{\em WISE}}
\newcommand\sdss{{SDSS}}

\newcommand\xmm{{\em XMM-Newton}}
\newcommand\nustar{{\em NuSTAR}}
\newcommand\xspec{{\sc xspec}}
\newcommand\oiii{[O\,{\sc iii}]}
\newcommand\av{$A_{\rm V}$}

\shorttitle{X-ray Observations of an OQQ}
\shortauthors{Greenwell et al.}
\graphicspath{{./}{figures/}}

\begin{document}

\title{\xmm\ and \nustar\ Observations of an Optically Quiescent Quasar}

\correspondingauthor{Claire Greenwell}
\email{cgreenwell11@gmail.com}

\author[0000-0002-7719-5809]{Claire Greenwell}
\affiliation{School of Physics \& Astronomy, University of Southampton, Highfield, Southampton SO17 1BJ, UK}
\affiliation{European Southern Observatory, Karl-Schwarzschild-Strasse 2, D-85748, Germany}

\author{Poshak Gandhi}
\affiliation{School of Physics \& Astronomy, University of Southampton, Highfield, Southampton SO17 1BJ, UK}

\author{George Lansbury}
\affiliation{European Southern Observatory, Karl-Schwarzschild-Strasse 2, D-85748, Germany}

\author{Peter Boorman}
\affiliation{Astronomical Institute, Academy of Sciences, Boční II 1401, CZ-14131 Prague, Czechia}
\affiliation{School of Physics \& Astronomy, University of Southampton, Highfield, Southampton SO17 1BJ, UK}

\author{Vincenzo Mainieri}
\affiliation{European Southern Observatory, Karl-Schwarzschild-Strasse 2, D-85748, Germany}

\author[0000-0003-2686-9241]{Daniel Stern}
\affiliation{Jet Propulsion Laboratory, California Institute of Technology, 4800 Oak Grove Drive, Mail Stop 169-221, Pasadena, CA 91109, USA}

\begin{abstract}

Optically quiescent quasars (OQQs) represent a recently systematised class of infrared-luminous active galactic nuclei (AGN) which have galaxy-like optical continua. They may represent an interesting, brief phase in the AGN life cycle, e.g. either cocooned within high-covering-factor media or indicative of recent triggering, though their nature remains unclear. Here, we present the first targeted simultaneous X-ray observations of an OQQ, our previously identified prototype, SDSS J075139.06+402811.2 at $z$=0.587. The source is significantly detected over 0.5--16\,keV with {\em XMM-Newton} and {\em NuSTAR}, unambiguously confirming the presence of current accretion activity. Spectral modelling yields an intrinsic luminosity $L_{\rm 2-10\,keV}$\, $\approx$\,4.4$ \times$10$^{43}$\, erg\,s$^{-1}$, well within the AGN regime, but underluminous relative to its infrared power. It is lightly obscured, with log $N_{\rm H}$\,[cm$^{-2}$]\,$\approx$\,22.

\end{abstract}

\keywords{Active galactic nuclei (16), X-ray active galactic nuclei (2035), X-ray astronomy (1810), Surveys (1671), Infrared galaxies (790)}

\section{Introduction} \label{sec:intro}

Knowledge of the demographics of the AGN population is essential to our understanding of how they evolve and influence their host galaxies. The majority of AGN are affected by obscuration, which hides central AGN signatures \citep[e.g.,][]{ananna_accretion_2020}. This obscuration can take many forms, both stable \citep[resulting in diverse AGN classes, e.g. Seyfert 1, Seyfert 2; ][]{padovani_active_2017}, and transient \citep[some \lq changing\rq\ look AGN (CLAGN), e.g. ][]{risaliti_rapid_2005, ricci_IC751_2016}. CLAGN can also present with a change in intrinsic flux, likely due to change in accretion state \citep[e.g.,][]{stern_mid-ir_2018}. Selecting an unbiased sample of AGN across these properties is challenging \citep[e.g.,][]{hickox_obscured_2018, asmus_local_2020}, and in this work we address a gap in current samples which could represent an interesting and unusual phase of obscuration.

AGN selection in each portion of the electromagnetic spectrum has both advantages and drawbacks \citep{brandt_cosmic_2015, padovani_active_2017}. Optical selection, using typical AGN emission lines, can miss heavily obscured or intrinsically faint objects \citep{hickox_obscured_2018}. Infrared (IR) selection offers the chance to investigate AGN without optical signatures -- the majority of IR emission from AGN is reprocessed in the dusty regions, and therefore relatively unbiased \citep{gandhi09, asmus_subarcsecond_2015}. Multi-wavelength studies using combinations of large surveys can be used to select AGN with interesting properties; furthermore extreme, unusual objects can be found in these vast datasets.

To systematically search for AGN with rare properties, we began an investigation based on bright, AGN-coloured IR sources from \wise\ \citep{wright_wide-field_2010} with no clear optical signatures of AGN presence -- specifically the \oiii\,$\lambda$5007\,\AA\ forbidden line, amongst the strongest lines found in the narrow-line regions of AGN. This optical--IR disparity would set any such AGN apart from standard selection techniques. We designate such sources as `Optically Quiescent Quasars (OQQs)'. Source powers were chosen to lie in the quasar regime in order to mitigate host galaxy dilution \citep{moran_hidden_2002} as a possible cause of the lack of optical emission lines. An in-depth study of a single, prototypical OQQ was presented in \citet{greenwell_candidate_2021}, and this paper demonstrates how X-ray observations provide an important tool for analysing the intrinsic emission of AGN -- particularly in the case of AGN with atypical properties. The results from our full survey will be discussed in detail in Greenwell et al. (in prep.).

Two likely physical scenarios that may explain the observed properties of OQQs are:

\begin{itemize}
    \item \lq Cocooned\rq\ AGN -- the optical emission lines are not seen because the AGN is completely enshrouded in a (presumably transient) \lq cocoon\rq\ of gas and dust. 
    \item \lq Young\rq\ AGN -- the AGN has recently switched on, and has not yet ionised the narrow line region (NLR): no \oiii\ line has yet been excited.
\end{itemize}

Both scenarios are interesting from an evolutionary perspective. AGN growth within fully enshrouding cocoons is suggested by some models \citep{fabian_obscured_1999}, and OQQs would represent a systematic search for this class of source, though such candidates appear in various prior sub-samples \citep[e.g., ][]{gandhi_chandrasources_2002, hviding_characterizing_2018}. Similarly, in the young AGN scenario, it may be possible to constrain the duty cycle of NLR excitation \citep{schawinski_active_2015, gezari_iPTF_2017}. However, the foremost requirement is an independent confirmation of the AGN nature of OQQs. This is especially important in the absence of optical AGN spectral signatures, and this is what we present herein. 

\section{Data} \label{sec:data}

In order to (a) confirm the presence of an AGN, and (b) constrain its spectral properties, the optimal tracer is X-ray emission. Luminous, nuclear X-ray radiation is unlikely to originate from any source other than an AGN -- intense star formation may produce (weaker) X-rays, but given the optically quiescent nature of OQQs \citep[$<$0.4 $M_\odot$ per year, stellar mass $\sim$ 10$^{11} M_\odot$; ][]{greenwell_candidate_2021}, sufficient star formation processes are not likely. \xmm\ provides good angular resolution and sensitivity, ideal to examine the soft X-rays. \nustar\ \citep{harrison_nuclear_2013} looks at the harder X-rays, allowing us to measure the intrinsic X-ray luminosity if SDSS~J075139.06+402811.2 (hereafter OQQ~J0751+4028) proves to be heavily obscured. Based on the IR--X-ray relationship \citep{asmus_subarcsecond_2015, stern_x-ray_2015} we predict that OQQ~J0751+4028 should be easily detected (under the assumption that it is indeed an AGN, and not heavily Compton-thick, at $z$=0.587, L$_{2-10 \textrm{ keV}}^{\rm predicted}\sim$2.6 $\times$ 10$^{44}$ erg\,s$^{-1}$), and analysis of its properties should be possible.

This analysis uses the following coordinated observations:
\begin{itemize}
    \item \nustar\ OBSID 60701009002: 50.6 ks, 2021 September 25 (start time: 12:06:09)
    \item \xmm\ OBSID 0884080101: 36.9 ks\footnote{After cleaning 16.4/28.3/29.3 ks on pn/MOS1/MOS2.} of exposure, 2021 September 25 (start time: 13:28:37)
\end{itemize}

\begin{figure}
    \centering
    \includegraphics[width=\columnwidth]{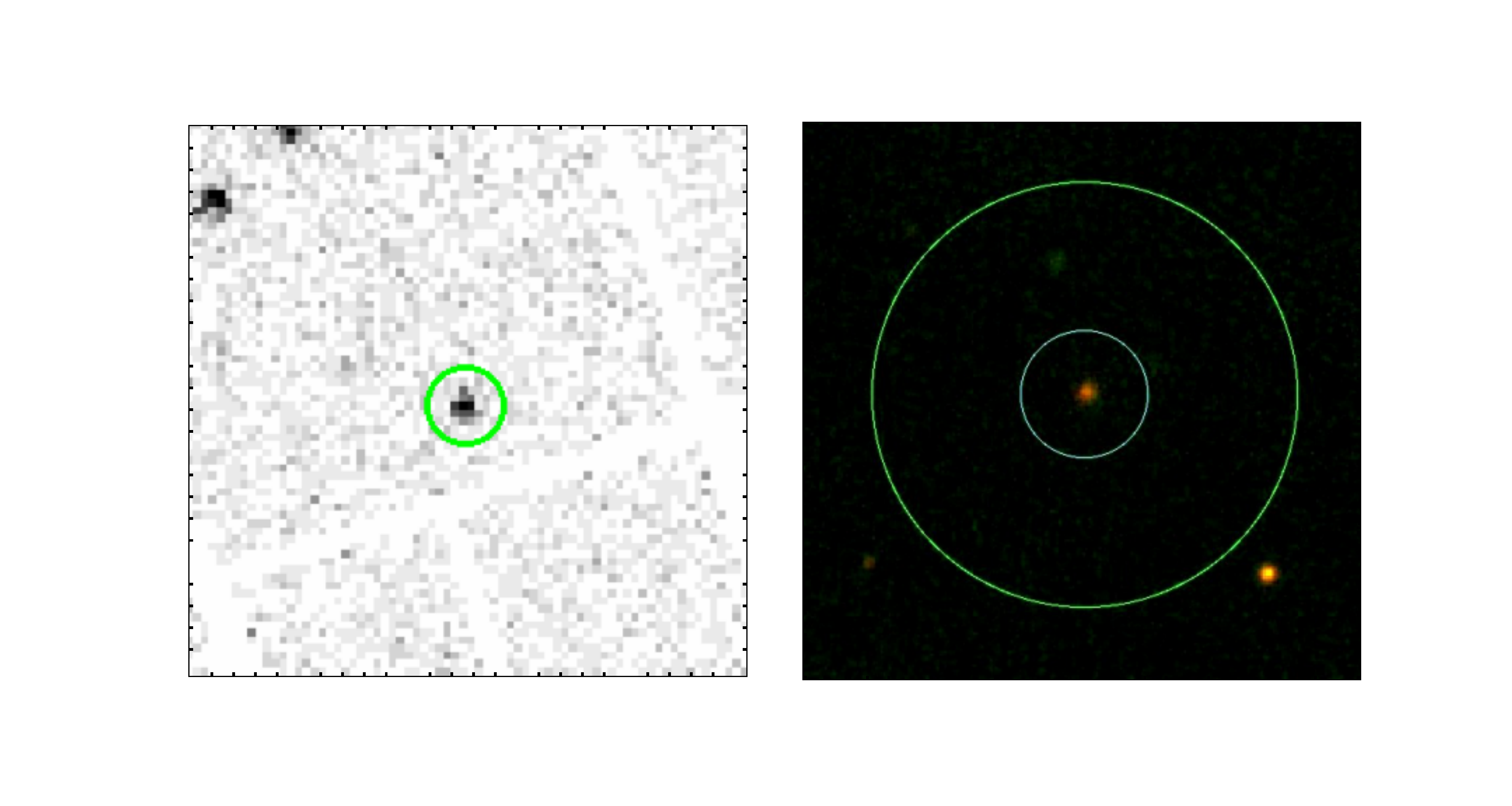}
    \vspace*{-1cm}
    \caption{(left) {\em XMM-Newton} pn image; (right) PanSTARRS {\it irg} bands converted to RGB image. Green circles have 20 arcsec radius, cyan circle has 6 arcsec radius; all are centred on the optical source coordinates.}
    \label{fig:images}
\end{figure}

The data were reduced using standard recommended selection criteria, including removal of appropriate background, using the \xmm\ Science Analysis Software\footnote{http://xmm.esa.int/sas/} and HEASoft\footnote{https://heasarc.gsfc.nasa.gov/docs/nustar/analysis/}. The target was detected significantly with \xmm\ pn, MOS1 and MOS2\footnote{0.5-10~keV; net counts 87/36/65}, and with \nustar\ FPMA\footnote{3-16~keV; net counts 55. Not detected in FPMB alone, due to higher background flux.}. Source extraction regions were circles with radii 45 and 20 arcsec for \nustar\ and \xmm\, respectively. Background regions were annuli with inner radius 100 arcsec, outer radius 180 arcsec (partially cutout to avoid a chip edge) and circles of radius 90 arcsec for \nustar\ and \xmm\, respectively. In optical data (PanSTARRS) it appears small and red, with no visible morphological disturbances (see Figure\,\ref{fig:images}).

\section{Methods} \label{sec:methods}

The data were fit within \xspec\ \citep{arnaud_xspec_1996}, v.12.12.0 with several models covering various types and structures of obscuration, with different levels of complexity. Relevant parameters were allowed to vary freely (although tied between datasets): normalisation, $\Gamma$ and $N_{\rm H}$.

Under the assumption that OQQ~J0751+4028 is an obscured AGN, the most basic combination of models we might expect to make a reasonable fit to the data is an absorbed powerlaw (photoelectric absorption at the source redshift and Compton scattering attenuating an intrinsic powerlaw; see top left panel in Figure \ref{fig:spectra_ratios}). This model can produce an acceptable fit to the data, with light absorption (see Table \ref{tab:fitresults}). However, the best fit $\Gamma$=1.08$\pm$0.16 is unusually hard for an AGN compared to typical intrinsic spectral indices of $\Gamma\sim$1.9 \citep{ricci_bat_2017}. There is some degeneracy between $N_{\rm H}$ and $\Gamma$ values, so in order to investigate the likelihood of a more typical intrinsic AGN power law, we fit the same model, but with a fixed $\Gamma$=1.9. This also produces an acceptable fit, with a slight increase in log($N_{\rm H}$; cm$^{-2}$)=21.56 to 22.30 (see Table \ref{tab:fitresults}).

Given the consistently hard $\Gamma$ seen in the absorbed powerlaw model we next investigate two more physically realistic models, still relatively simple: torus reprocessing \citep[{\sc mytorus} in coupled mode with covering factor fixed at 0.5; ][]{murphy_x-ray_2009}; and the spherical obscuration \citep[{\sc trans} model from {\sc BNtorus};][hereafter {\sc BNsphere}]{brightman_xmm-newton_2011}.

{\sc BNsphere} represents the physically expected obscuring structure around a \lq Cocooned\rq\ AGN. {\sc mytorus} does not specifically allow full covering, and is restricted to $\Gamma \ge$1.4, higher than previously found $\Gamma$ values. {\sc BNsphere} is also limited, but to $\Gamma \ge$ 1.0. 

Other parameters are fixed to simplify the modelling. {\sc mytorus} inclination was set to 90~$\deg$ (i.e., through the torus, to fit with our assumption of an obscured AGN). {\sc BNsphere} iron and total elemental abundances were set to solar values.

The results for the models thus far consistently produce a hard $\Gamma$ ($\le$ 1.5). This would be an unusual intrinsic value for an AGN, and could indicate that we are underestimating the optical depth of obscuration present in the system, a possibility also hinted at by the lower than expected luminosity. None of the models above can produce a log($N_{\rm H}$) greater than $\sim$22.3, thus we propose an alternative: a thick spherical obscurer (represented in {\sc xspec} with {\sc BNsphere}) in tandem with a scattered fraction ({\sc constant}) of the intrinsic powerlaw ({\sc zcutoffpl}). The scattered fraction represents a \lq leak\rq\ through Compton thin obscuration from an otherwise Compton thick sphere. With this we can investigate higher $N_{\rm H}$ values while still providing a satisfactory fit to the softer X-rays, i.e. dominated by the scattered powerlaw rather than a Compton hump. Here we show the results with scattering fraction fixed at 12\%, which produces an intrinsic X-ray luminosity close to that predicted from the 12\,\micron\ luminosity.

Fixing the value of $\Gamma$ makes a firm assumption about the nature of the intrinsic AGN emission, which we can make less stringent with Bayesian X-ray Analysis (BXA) \citep{buchner_x-ray_2014}. We can include knowledge about the likely physical characteristics by selecting an appropriate prior: a Gaussian prior for $\Gamma$\footnote{$N_{\rm H}$ and normalisation have log uniform priors.} is appropriate as it allows a physically motivated preference towards likely values, and excludes unphysical values. We try (a) a loose Gaussian prior: $\Gamma$=2.0$\pm$0.3; and (b) a stricter Gaussian prior $\Gamma$=1.9$\pm$0.1. As shown in Table \ref{tab:fitresults}, these both produce an acceptable fit to the data but struggle to produce a $\Gamma$ value not unusually hard for an AGN. We can use the Bayesian evidence to compare the models, and therefore determine which is most likely to have produced the observed data.

The final set of models we compare is: (a) {\sc cabs*zwabs*pow}: with uniform, single value and Gaussian priors on $\Gamma$; (b) {\sc mytorus}; (c) {\sc BNsphere}\footnote{{\sc BNtorus} is known to have inaccuracies in the reflected component \citep[e.g.][]{balokovic_new_2018}; here we only consider the spherical component with no opening angle which should not be affected.}; and (d) \lq leaky sphere\rq.

\begin{deluxetable*}{cccccccc}
\label{tab:fitresults}
\tablenum{1}
\tablecaption{Spectral modelling results. Column details: (1) {\sc xspec} model; (2) $\Gamma$ prior; (3) $\Gamma$; (4) $N_{\rm H}$; (5) \lq sphere\rq\ $N_{\rm H}$; (6) unabsorbed 2-10~keV luminosity; (7) fit statistic; (8) Bayes factor compared to {\sc BNsphere}.}
\tablewidth{0pt}
\tablehead{
\colhead{Model} & \colhead{$\Gamma$ prior} & \colhead{$\Gamma$} & \colhead{$\log N_{\rm H}$} & \colhead{$\log N_{\rm H, sphere}$} & \colhead{$\log L_{2-10 \textrm{ keV}}$} & \colhead{cstat/d.o.f.} & \colhead{Bayes factor} \\
\colhead{ } & \colhead{ } & \colhead{ } & \colhead{(cm$^{-2}$)} & \colhead{(cm$^{-2}$)} & \colhead{(erg s$^{-1}$)} & \colhead{ } & \colhead{(normalised)}
}
\decimalcolnumbers
\startdata
{\sc cabs*zwabs*pow} & Uniform     & 0.98$_{-0.09}^{+0.18}$ & 21.62$_{-0.44}^{+0.27}$ & & 43.64$\pm$0.04 & 334.1/377 & --   \\
{\sc cabs*zwabs*pow} & Fixed 1.9   & 1.90                   & 22.30$_{-0.11}^{+0.08}$ & & 43.68$\pm$0.05 & 346.2/378 & 31.6 \\
{\sc cabs*zwabs*pow} & 1.9$\pm$0.1 & 1.79$_{-0.10}^{+0.09}$ & 22.26$_{-0.13}^{+0.08}$ & & 43.67$\pm$0.05 & 337.7/377 & 12.6 \\
{\sc cabs*zwabs*pow} & 2.0$\pm$0.3 & 1.26$_{-0.16}^{+0.25}$ & 22.03$_{-0.41}^{+0.15}$ & & 43.64$\pm$0.05 & 334.1/377 & 1.6  \\
{\sc mytorus}        & 2.0$\pm$0.3 & 1.47$_{-0.08}^{+0.17}$ & 22.05$_{-0.00}^{+0.12}$ & & 43.65$\pm$0.04 & 336.6/377 & 2.0  \\
{\sc BNsphere}       & 2.0$\pm$0.3 & 1.32$_{-0.19}^{+0.21}$ & 21.95$_{-0.36}^{+0.15}$ & & 43.64$\pm$0.05 & 334.1/377 & best ($\equiv$1.0) \\
\lq leaky sphere\rq, 12\% & 2.0$\pm$0.3 & 2.19$_{-0.22}^{-0.18}$ & 22.26$_{-0.17}^{-0.10}$ & 24.08$_{-0.14}^{-0.14}$ & 44.45$\pm$0.09 & 341.9/376 & 6.3 \\
\enddata
\end{deluxetable*}

\section{Results} \label{sec:results}

\begin{figure*}
    \centering
    \includegraphics[width=0.9\textwidth]{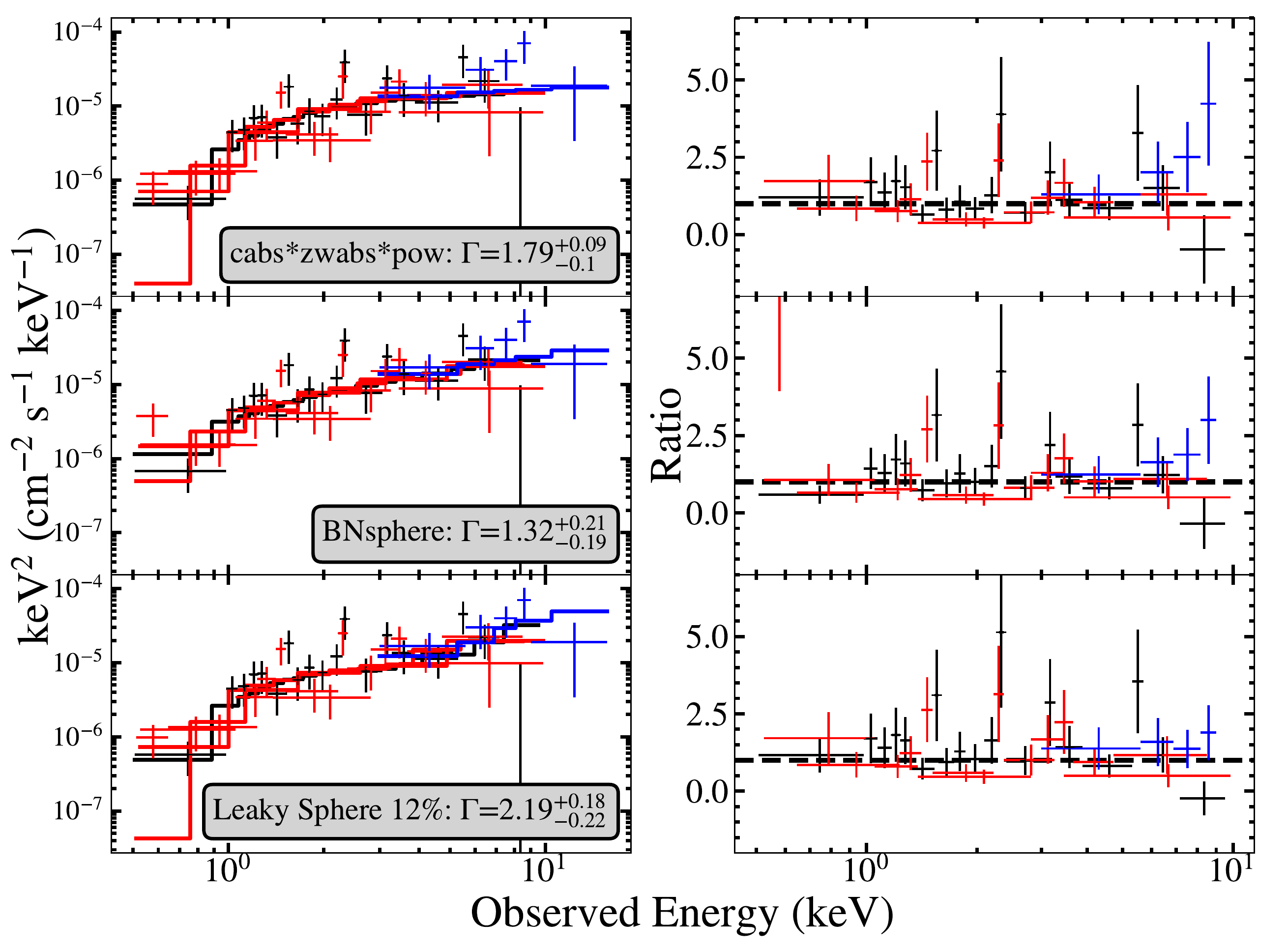}
    \caption{Spectra and ratio between data and model for (top) {\sc cabs*zwabs*pow}, $\Gamma$=1.79; (middle) {\sc BNsphere}, $\Gamma$=1.35; (bottom) \lq leaky sphere\rq, $\Gamma$=2.19. Data sShown is \xmm-pn (black), \xmm-MOS (red) and \nustar\ FPMA (blue), binned to a minimum of 2 counts per bin.}
    \label{fig:spectra_ratios}
\end{figure*}

Across all models, the results show that the data can be explained by an obscured AGN. The parameters vary, and in some cases may indicate unusual values, but all are consistent with the presence of an AGN. Full results for all models are shown in Table \ref{tab:fitresults}. None of the first six models show significant residuals, except possibly towards the hard end, and all are reasonable fits to the data (see Figure \ref{fig:spectra_ratios}).

The preferred solution according to the Bayes factors is {\sc BNsphere} (Table \ref{tab:fitresults})\footnote{{\sc cabs*zwabs*pow} with uniform prior fit is discarded from this point; the lack of constraint allows values of $\Gamma$ that are not physically likely.}. Comparison of (a) {\sc BNsphere} and (b) \lq leaky sphere\rq\ (scattering fraction 12\%) shows that LS is not favoured. However, it is also not strongly counter-indicated (i.e. the relative Bayes factor is low\footnote{A value of 6.3 is \lq substantial evidence\rq\ in favour of {\sc BNsphere}, but not decisive, according to the Jeffreys scale \citep{buchner_x-ray_2014}.}).

\begin{figure*}
    \centering
    \includegraphics[width=0.9\textwidth]{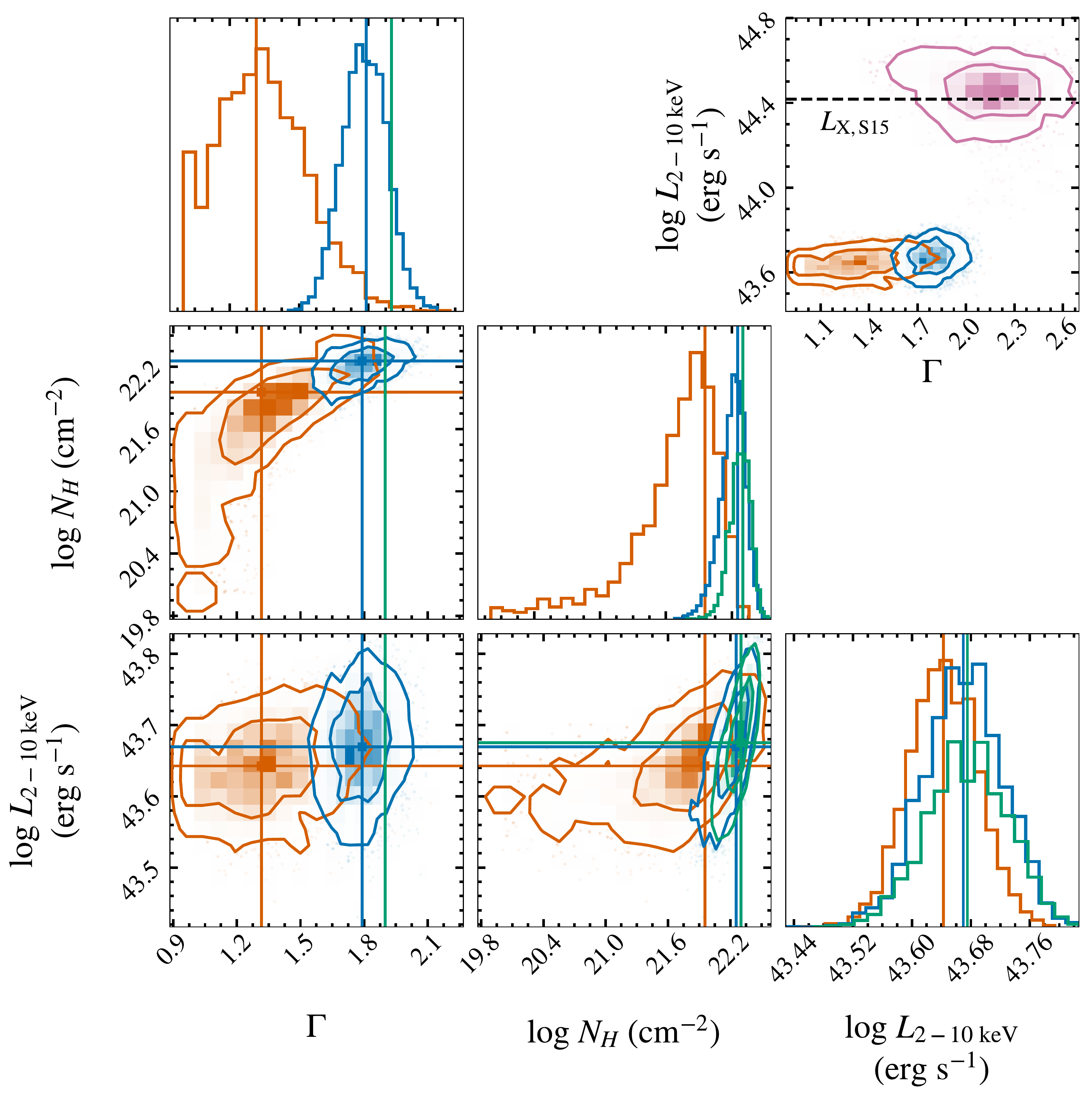}
    \caption{Corner plot of results with: (a) {\sc cabs*zwabs*pow}, restrictive prior on $\Gamma$ (blue); (b) {\sc BNsphere}, physically representative prior on $\Gamma$ (orange), (c) {\sc cabs*zwabs*pow}, fixed $\Gamma$=1.9 (green). Best fit values for each parameter are shown with a solid line. (top right) $L_{2-10}$ vs. $\Gamma$ contour with the IR-predicted luminosity shown \citep[][dashed line]{stern_x-ray_2015} -- the luminosity contours for the lightly obscured models are decisively below this level. Also shown are the results for the 12\% scattered \lq leaky sphere\rq, which is closer to predicted (pink).}
    \vspace{2mm}
    \label{fig:corner}
\end{figure*}

{\sc BNsphere} returns a preferred $\Gamma$=1.32$^{+0.21}_{-0.19}$ --- unusually hard but not impossible for an AGN \citep[e.g.; ][]{ricci_bat_2017}  --- and log ($N_{\rm H}$ / cm$^{-2}$) = 21.95 (see Figure\,\ref{fig:corner}). Forcing $\Gamma$ to a more typical AGN value of 1.9 increases the log($N_{\rm H}$ / cm$^{-2}$) slightly to 22.3, but does not significantly affect the intrinsic luminosity (see Figure \ref{fig:corner}). The 12\% \lq leaky sphere\rq\ model produces a softer $\Gamma$ and, if closer to the truth, may imply a luminosity much closer to expected.

That {\sc BNsphere} is a reasonable fit to the data implies that the \lq cocooned AGN\rq\ scenario may be a feasible explanation for the observed properties.  Obscuring material relatively close to the AGN may prevent ionizing radiation from within reaching farther out, and thus no narrow \oiii\ emission. We can also consider a situation where the NLR exists within a cocoon on intermediate scales between the torus and the inner galaxy (perhaps $\sim$\,tens of pc), just large enough to cocoon the inner NLRDepending on the gas--to--dust ratio in the obscuring material, relatively thin columns may be sufficient to extinct any \oiii. Based on the \oiii\ deficit between the general OQQ population and QSO2s (Greenwell et al., in prep.), a median \av\,=\,4.7\,mag is sufficient to extinct the theoretical \oiii\ line. This is equivalent to an NLR obscuring neutral gas column density of log($N_{\rm H}$/cm$^{-2}$)$\sim$\,22.8 \citep[assuming gas--to--dust ratios in AGN environments from 
][]{maiolino_dust_2001}. This is higher than found in this analysis; however, the NLR--obscuring column is distinct from the line of sight X-ray--obscuring column, which may explain this difference.

The Fe K$\alpha$ line (rest energy 6.4~keV) observed in many AGN spectra \citep[e.g.][]{nandra_on_2006} originates from reprocessing of AGN emission in optically thick obscuring matter. For lightly obscured AGN, \citet{shu_cores_2010} find a relationship between detected narrow Fe K$\alpha$ EW and unabsorbed 2-10~keV luminosity\footnote{With large scatter, and from higher resolution {\it Chandra} data.}, which we can use to roughly estimate an expected Fe K$\alpha$ EW of 45\,eV. We place an upper limit on the equivalent width (EW) of a putative narrow line by adding an unresolved Gaussian component to our transmission spectrum, with a width of 0.1~keV, finding an EW$\lesssim$26 eV -- a low value that implies no line is likely to be present, further reinforcing the conclusion that OQQ~J0751+4028 is only lightly obscured.

\begin{figure*}
    \centering
    \includegraphics[width=0.9\textwidth]{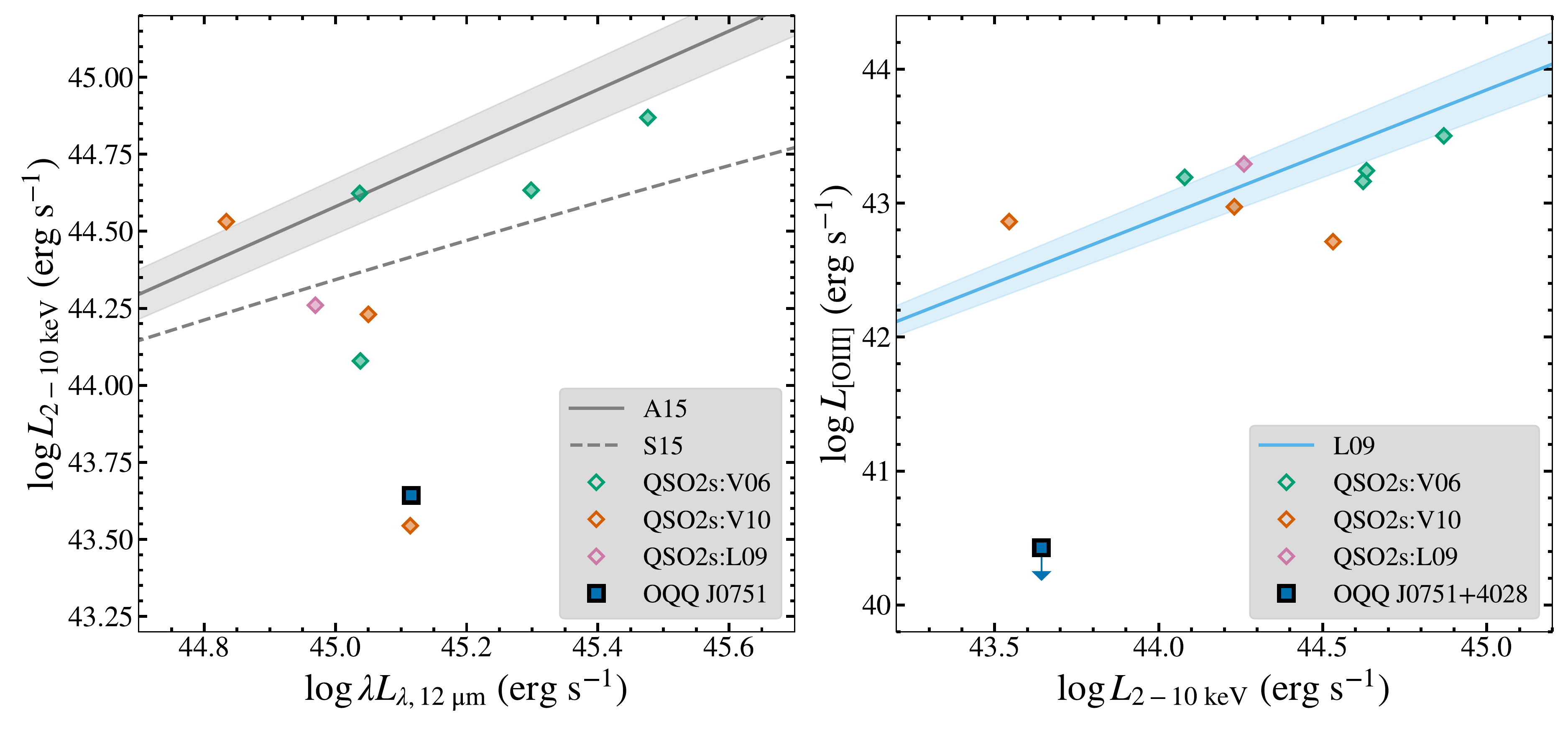}
    \caption{A comparison between OQQ~J0751+4028 and QSO2s: (left) unabsorbed 2-10~keV luminosity against 12 \micron\ luminosity. X-ray luminosities for QSO2s are from \citet{vignali_quest_2006} (V06; green diamonds), \citet{vignali_disovery_2010} (V10; orange diamonds), and \citet{lamastra_bolometric_2009} (L09; pink diamonds). Relationships from \citet[][A15]{asmus_subarcsecond_2015}, and \citet[][S15; 12\,\micron\ luminosities were converted to 6\micron\ luminosities using a relationship derived from the QSO template of \citet{hao_distribution_2007}]{stern_x-ray_2015}, are shown as grey lines. (right) unabsorbed 2-10~keV luminosity against \oiii\ luminosity, with the relationship from \citet[][L09]{lamastra_bolometric_2009}.}
    \label{fig:L12_oiii_xray}
\end{figure*}

The intrinsic rest frame 2-10~keV luminosity (according to {\sc BNsphere}) is 4.39 $\times$ 10$^{43}$ erg\,s$^{-1}$. The IR luminosity (1.30 $\times$ 10$^{45}$\,erg\,s$^{-1}$ at 12\,\micron) of OQQ~J0751+4028 implies a 2-10\,keV luminosity of 2.61 $\times$ 10$^{44}$\,erg\,s$^{-1}$ \citep[from the 6\,\micron/2-10\,keV relation;][]{stern_x-ray_2015}. Comparing these values, we find that the unabsorbed luminosity is $\sim$6 times lower than expected, but still easily above the threshold of 10$^{42-43}$ erg\,s$^{-1}$ generally accepted for an AGN. In Figure \ref{fig:L12_oiii_xray} we compare the properties of OQQ~J0751+4028 against a sample of Type 2 QSOs selected from \sdss\ with significant \oiii\ emission \citep[QSO2s; ][]{reyes_space_2008, yuan_spectroscopic_2016}. Figure \ref{fig:L12_oiii_xray} shows that OQQ~J0751+4028 lies below the IR-predicted values from \citet{asmus_subarcsecond_2015} or \citet{stern_x-ray_2015}, and also below the majority of QSO2s, which tend to fall closer to their predicted values. Figure \ref{fig:L12_oiii_xray} (right hand panel) shows the measured X-ray luminosities of the QSO2s, along with the reported \oiii/X-ray relationship from \citet{lamastra_bolometric_2009} -- the majority of QSO2s lie close to the empirical prediction, but OQQ~J0751+4028 is far from typical. Conversely, if we consider the \lq leaky sphere\rq\ model, the intrinsic X-ray luminosity may be closer to IR-predicted expectations (see Table \ref{tab:fitresults}); however, it would be further offset from the \oiii/X-ray relationship.

\section{Discussion}\label{sec:discussion}

One important aim of this work is to place OQQ~J0751+4028 (and in the future, its fellow OQQs) into context with the wider ranks of AGN. We begin by considering what the results from this X-ray study allow us to infer about the intrinsic nature of this object. Crucially, the absorption-corrected luminosity shows that OQQ~J0751+4028 is an AGN, regardless of the specific best-fit model that we adopt. This confirms the presence of ongoing accretion activity, and gives credence to our IR selection, despite the apparent optical quiescence.

OQQ~J0751+4028 must be obscured in X-rays, but is less likely to be Compton thick than thin. The Bayes factor for the highly obscured \lq leaky sphere\rq\ compared to the lightly obscured {\sc BNsphere} is $\sim$6: a less likely fit but not decisively so. Recalling from the introduction our suggested scenarios regarding the nature of an optically quiescent AGN (\lq cocooned\rq\ and \lq young\rq\ AGN), we can only conclude at this stage that both remain possible. Nuclear optical extinction is also required in both scenarios in order to hide the broad line region, which we do not see.

The light obscuration we observe could be a more physically likely possibility for enshrouding material than optically thick material, supporting the idea of a \lq cocooned\rq\ AGN, or similar to larger-scale host obscuration \citep[e.g.,][]{buchner_galaxy_2017}. This interpretation could be consistent with a low intrinsic X-ray luminosity AGN if the Eddington ratio is low, as higher intensity AGN are associated with lower covering factors \citep[e.g.,][]{ricci_close_2017}.

An AGN in the process of switching on (a \lq young\rq\ AGN) might also show weak X-ray emission as it transitions to full accretion power. \citet{kollatschny_optical_2020} present the opposite case -- a {\it switching off} AGN -- in which they see a dramatic decrease in observed X-ray luminosity concurrent with a change in type from Seyfert 1 to 1.9 (i.e. a reduction in the broad emission lines, but still with clear narrow lines). They find no evidence that this change is caused by absorption, indicating that it is an intrinsic luminosity change. Reversing this, we might expect to see the X-ray increase before the appearance of narrow lines.

Finally, we must consider the possibility that OQQ~J0751+4028 is a fully \lq mature\rq\ AGN {\it intrinsically} lacking \oiii. Analysis of relationships between various emission-line properties of AGN has shown that many of these properties correlate \citep[Eigenvector 1;][]{boroson_emission-line_1992}. \citet{shen_diversity_2014} show that the observed anti-correlation between \oiii\ and the relative strength of Fe {\sc II} emission can be explained by changes in Eddington ratio; an AGN seen to lack \oiii\ could then be explained by a very high accretion rate. The higher $\Gamma$ and intrinsic luminosities of the \lq Leaky Sphere\rq\ model may suggest this to be the case, however a low \lq leak fraction\rq\ would be required and the fit statistics indicate this is unlikely. The stellar mass of OQQ~J0751+4028, while uncertain, implies a high BH mass \citep[$\sim 5 \times 10^8 M_\odot$; ][]{kormendy_coevolution_2013}, and consequently a very high luminosity to reach Eddington accretion levels -- higher than seen in the \wise\ measurements.

OQQ-like objects are not new. For example, other groups of AGN that have notable similarities to OQQs include weak line quasars (WLQs) with weak X-ray emission and weak or absent emission lines \citep[e.g.,][]{wu_x-ray_2012}, and X-ray bright, optically normal galaxies (XBONGs) selected as AGN in X-rays, but showing no optical AGN signatures. In some cases, this is due to dilution by bright host galaxies of lower luminosity AGN \citep{moran_hidden_2002} -- in contrast to the OQQs, where AGN emission in the IR is bright, and dominates over their hosts.

Our OQQ selection is a first attempt to systematically search for this class of optically quiescent AGN and our work unambiguously establishes that some AGN can present bright ongoing nuclear accretion activity in X-rays, yet show no signs of this in the optical. Details of our full sample will be presented in Greenwell et al. (in prep.).

\section{Summary}\label{sec:summary}

We present the first targeted, simultaneous, hard and soft X-ray observations of an optically quiescent quasar, demonstrating \textit{unequivocally} that OQQ~J0751+4028 is an AGN. It is lightly obscured ($N_{\rm H}$ $\approx$\,10$^{22}$ cm$^{-2}$) and, while bright enough to confirm the presence of an AGN (L$_{2-10 \textrm{ keV}}$=4.37 $\times$ 10$^{43}$ erg\,s$^{-1}$), our most likely model suggests it is less X-ray luminous than would be expected given its IR properties, and in comparison to an \oiii-bright population. This result shows conclusively that the OQQ selection technique introduced in \citet{greenwell_candidate_2021}, and discussed in depth in Greenwell et al. (in prep.), can uncover previously unknown AGN.

\bigskip

\begin{acknowledgments}
We thank the referee for their helpful comments and suggestions. This research is funded by UKRI. CG is supported by a University of Southampton Mayflower studentship. PG acknowledges support from STFC and a UGC-UKIERI Thematic partnership (STFC grant number ST/V001000/1). PGB acknowledges financial support from the Czech Science Foundation project No. 22-22643S. This publication makes use of data products from \wise\, which is a project of University of California, Los Angeles, and the Jet Propulsion Laboratory (JPL)/California Institute of Technology (Caltech), funded by the National Aeronautics and Space Administration (NASA). Funding for \sdss-III has been provided by the Alfred P. Sloan Foundation, the Participating Institutions, the National Science Foundation, and the U.S. Department of Energy Office of Science. The \sdss-III web site is http://www.sdss3.org/. This work made use of observations obtained with {\it XMM-Newton}, an ESA science mission with instruments and contributions directly funded by ESA Member States and NASA. This research made use of data from the {\it NuSTAR} mission, a project led by Caltech, managed by JPL, and funded by NASA. This research has made use of the {\it NuSTAR} Data Analysis Software (NuSTARDAS) jointly developed by the ASI Science Data Center (ASDC, Italy) and the Caltech (USA). The NASA/IPAC Infrared Science Archive (IRSA) operated by JPL under contract with NASA was used. This research has made use of data obtained through the High Energy Astrophysics Science Archive Research Center Online Service, provided by the NASA/Goddard Space Flight Center.

\end{acknowledgments}

\vspace{5mm}
\facilities{\xmm, \nustar, \sdss, \wise}

\software{astropy \citep{2013A&A...558A..33A,2018AJ....156..123A},
            xspec \citep{arnaud_xspec_1996},
            corner \citep{corner},
            BXA \citep{buchner_x-ray_2014},
            UltraNest \citep{buchner_ultranest_2021}
          }

\appendix

\section{\xspec\ Models}

The models used in \xspec\ are as follows:
\begin{itemize}
    \item Absorbed powerlaw: {\sc constant*phabs*cabs*zwabs*pow}
    \item {\sc mytorus}: {\sc constant*phabs(zpowerlw*etable\{mytorus\_Ezero\_v00.fits\} + \\constant*atable\{mytorus\_scatteredH500\_v00.fits\})}
    \item {\sc BNsphere}: {\sc constant*phabs*atable\{sphere0708.fits\}}
    \item \lq leaky sphere\rq: {\sc constant*phabs*(atable\{sphere0708.fits\}+constant*zwabs*zcutoffpl)}
    \item Absorbed powerlaw with neutral iron line: {\sc constant*phabs(cabs*zwabs*powerlaw+zgauss)}
\end{itemize}

All models include Galactic absorption ({\sc phabs}) of $N_{\rm H}$=5.6 $\times$ 10$^{20}$ cm$^{-2}$ \citep{HI4PI_collab_2016}. A cross calibration constant of 0.93:1.02:0.98:1.00 \citep[e.g.][]{madsen_calibration_2015} was used for \xmm(pn):\xmm(MOS1):\xmm(MOS2):\nustar; this was fixed because the observations were simultaneous. Spectra were binned to 3/1 counts per bin for \nustar/\xmm, and fit with {\tt wstat} \citep{wachter_parameter_1979} -- the version of {\tt cstat} \citep{cash_parameter_1979} used by \xspec\ when a background is included: \url{https://heasarc.gsfc.nasa.gov/xanadu/xspec/manual/node318.html\#AppendixStatistics.}.

\section{Data Availability}

The optical and IR data underlying this article are publicly available from the \textit{WISE} All-Sky Survey (DOI 10.26131/IRSA1) and SDSS DR15. X-ray data are currently in the proprietary period, but will become publically available with the Observation IDs listed above.

\bibliography{lib}{}

\begin{thebibliography}{}
\expandafter\ifx\csname natexlab\endcsname\relax\def\natexlab#1{#1}\fi
\providecommand{\url}[1]{\href{#1}{#1}}
\providecommand{\dodoi}[1]{doi:~\href{http://doi.org/#1}{\nolinkurl{#1}}}
\providecommand{\doeprint}[1]{\href{http://ascl.net/#1}{\nolinkurl{http://ascl.net/#1}}}
\providecommand{\doarXiv}[1]{\href{https://arxiv.org/abs/#1}{\nolinkurl{https://arxiv.org/abs/#1}}}

\bibitem[{{Ananna} {et~al.}(2020){Ananna}, {Treister}, {Urry}, {Ricci},
  {Hickox}, {Padmanabhan}, {Marchesi}, \&
  {Kirkpatrick}}]{ananna_accretion_2020}
{Ananna}, T.~T., {Treister}, E., {Urry}, C.~M., {et~al.} 2020, \apj, 889, 17,
  \dodoi{10.3847/1538-4357/ab5aef}

\bibitem[{{Arnaud}(1996)}]{arnaud_xspec_1996}
{Arnaud}, K.~A. 1996, in Astronomical Society of the Pacific Conference Series,
  Vol. 101, Astronomical Data Analysis Software and Systems V, ed. G.~H.
  {Jacoby} \& J.~{Barnes}, 17

\bibitem[{Asmus {et~al.}(2015)Asmus, Gandhi, Hönig, Smette, \&
  Duschl}]{asmus_subarcsecond_2015}
Asmus, D., Gandhi, P., Hönig, S.~F., Smette, A., \& Duschl, W.~J. 2015,
  \mnras, 454, 766, \dodoi{10.1093/mnras/stv1950}

\bibitem[{{Asmus} {et~al.}(2020){Asmus}, {Greenwell}, {Gandhi}, {Boorman},
  {Aird}, {Alexander}, {Assef}, {Baldi}, {Davies}, {H{\"o}nig}, {Ricci},
  {Rosario}, {Salvato}, {Shankar}, \& {Stern}}]{asmus_local_2020}
{Asmus}, D., {Greenwell}, C.~L., {Gandhi}, P., {et~al.} 2020, \mnras, 494,
  1784, \dodoi{10.1093/mnras/staa766}

\bibitem[{{Astropy Collaboration} {et~al.}(2013){Astropy Collaboration},
  {Robitaille}, {Tollerud}, {Greenfield}, {Droettboom}, {Bray}, {Aldcroft},
  {Davis}, {Ginsburg}, {Price-Whelan}, {Kerzendorf}, {Conley}, {Crighton},
  {Barbary}, {Muna}, {Ferguson}, {Grollier}, {Parikh}, {Nair}, {Unther},
  {Deil}, {Woillez}, {Conseil}, {Kramer}, {Turner}, {Singer}, {Fox}, {Weaver},
  {Zabalza}, {Edwards}, {Azalee Bostroem}, {Burke}, {Casey}, {Crawford},
  {Dencheva}, {Ely}, {Jenness}, {Labrie}, {Lim}, {Pierfederici}, {Pontzen},
  {Ptak}, {Refsdal}, {Servillat}, \& {Streicher}}]{2013A&A...558A..33A}
{Astropy Collaboration}, {Robitaille}, T.~P., {Tollerud}, E.~J., {et~al.} 2013,
  \aap, 558, A33, \dodoi{10.1051/0004-6361/201322068}

\bibitem[{{Astropy Collaboration} {et~al.}(2018){Astropy Collaboration},
  {Price-Whelan}, {Sip{\H{o}}cz}, {G{\"u}nther}, {Lim}, {Crawford}, {Conseil},
  {Shupe}, {Craig}, {Dencheva}, {Ginsburg}, {VanderPlas}, {Bradley},
  {P{\'e}rez-Su{\'a}rez}, {de Val-Borro}, {Aldcroft}, {Cruz}, {Robitaille},
  {Tollerud}, {Ardelean}, {Babej}, {Bach}, {Bachetti}, {Bakanov}, {Bamford},
  {Barentsen}, {Barmby}, {Baumbach}, {Berry}, {Biscani}, {Boquien}, {Bostroem},
  {Bouma}, {Brammer}, {Bray}, {Breytenbach}, {Buddelmeijer}, {Burke},
  {Calderone}, {Cano Rodr{\'\i}guez}, {Cara}, {Cardoso}, {Cheedella}, {Copin},
  {Corrales}, {Crichton}, {D'Avella}, {Deil}, {Depagne}, {Dietrich}, {Donath},
  {Droettboom}, {Earl}, {Erben}, {Fabbro}, {Ferreira}, {Finethy}, {Fox},
  {Garrison}, {Gibbons}, {Goldstein}, {Gommers}, {Greco}, {Greenfield},
  {Groener}, {Grollier}, {Hagen}, {Hirst}, {Homeier}, {Horton}, {Hosseinzadeh},
  {Hu}, {Hunkeler}, {Ivezi{\'c}}, {Jain}, {Jenness}, {Kanarek}, {Kendrew},
  {Kern}, {Kerzendorf}, {Khvalko}, {King}, {Kirkby}, {Kulkarni}, {Kumar},
  {Lee}, {Lenz}, {Littlefair}, {Ma}, {Macleod}, {Mastropietro}, {McCully},
  {Montagnac}, {Morris}, {Mueller}, {Mumford}, {Muna}, {Murphy}, {Nelson},
  {Nguyen}, {Ninan}, {N{\"o}the}, {Ogaz}, {Oh}, {Parejko}, {Parley}, {Pascual},
  {Patil}, {Patil}, {Plunkett}, {Prochaska}, {Rastogi}, {Reddy Janga},
  {Sabater}, {Sakurikar}, {Seifert}, {Sherbert}, {Sherwood-Taylor}, {Shih},
  {Sick}, {Silbiger}, {Singanamalla}, {Singer}, {Sladen}, {Sooley},
  {Sornarajah}, {Streicher}, {Teuben}, {Thomas}, {Tremblay}, {Turner},
  {Terr{\'o}n}, {van Kerkwijk}, {de la Vega}, {Watkins}, {Weaver}, {Whitmore},
  {Woillez}, {Zabalza}, \& {Astropy Contributors}}]{2018AJ....156..123A}
{Astropy Collaboration}, {Price-Whelan}, A.~M., {Sip{\H{o}}cz}, B.~M., {et~al.}
  2018, \aj, 156, 123, \dodoi{10.3847/1538-3881/aabc4f}

\bibitem[{{Balokovi{\'c}} {et~al.}(2018){Balokovi{\'c}}, {Brightman},
  {Harrison}, {Comastri}, {Ricci}, {Buchner}, {Gandhi}, {Farrah}, \&
  {Stern}}]{balokovic_new_2018}
{Balokovi{\'c}}, M., {Brightman}, M., {Harrison}, F.~A., {et~al.} 2018, \apj,
  854, 42, \dodoi{10.3847/1538-4357/aaa7eb}

\bibitem[{{Boroson} \& {Green}(1992)}]{boroson_emission-line_1992}
{Boroson}, T.~A., \& {Green}, R.~F. 1992, \apjs, 80, 109,
  \dodoi{10.1086/191661}

\bibitem[{{Brandt} \& {Alexander}(2015)}]{brandt_cosmic_2015}
{Brandt}, W.~N., \& {Alexander}, D.~M. 2015, \aapr, 23, 1,
  \dodoi{10.1007/s00159-014-0081-z}

\bibitem[{{Brightman} \& {Nandra}(2011)}]{brightman_xmm-newton_2011}
{Brightman}, M., \& {Nandra}, K. 2011, \mnras, 413, 1206,
  \dodoi{10.1111/j.1365-2966.2011.18207.x}

\bibitem[{{Buchner}(2021)}]{buchner_ultranest_2021}
{Buchner}, J. 2021, J. Open Source Softw., 6, 3001, \dodoi{10.21105/joss.03001}

\bibitem[{{Buchner} \& {Bauer}(2017)}]{buchner_galaxy_2017}
{Buchner}, J., \& {Bauer}, F.~E. 2017, \mnras, 465, 4348,
  \dodoi{10.1093/mnras/stw2955}

\bibitem[{{Buchner} {et~al.}(2014){Buchner}, {Georgakakis}, {Nandra}, {Hsu},
  {Rangel}, {Brightman}, {Merloni}, {Salvato}, {Donley}, \&
  {Kocevski}}]{buchner_x-ray_2014}
{Buchner}, J., {Georgakakis}, A., {Nandra}, K., {et~al.} 2014, \aap, 564, A125,
  \dodoi{10.1051/0004-6361/201322971}

\bibitem[{{Cash}(1979)}]{cash_parameter_1979}
{Cash}, W. 1979, \apj, 228, 939, \dodoi{10.1086/156922}

\bibitem[{Fabian(1999)}]{fabian_obscured_1999}
Fabian, A.~C. 1999, \mnras, 308, L39, \dodoi{10.1046/j.1365-8711.1999.03017.x}

\bibitem[{Foreman-Mackey(2016)}]{corner}
Foreman-Mackey, D. 2016, J. Open Source Softw., 1, 24,
  \dodoi{10.21105/joss.00024}

\bibitem[{{Gandhi} {et~al.}(2002){Gandhi}, {Crawford}, \&
  {Fabian}}]{gandhi_chandrasources_2002}
{Gandhi}, P., {Crawford}, C.~S., \& {Fabian}, A.~C. 2002, \mnras, 337, 781,
  \dodoi{10.1046/j.1365-8711.2002.05805.x}

\bibitem[{{Gandhi} {et~al.}(2009){Gandhi}, {Horst}, {Smette}, {H{\"o}nig},
  {Comastri}, {Gilli}, {Vignali}, \& {Duschl}}]{gandhi09}
{Gandhi}, P., {Horst}, H., {Smette}, A., {et~al.} 2009, \aap, 502, 457,
  \dodoi{10.1051/0004-6361/200811368}

\bibitem[{{Gezari} {et~al.}(2017){Gezari}, {Hung}, {Cenko}, {Blagorodnova},
  {Yan}, {Kulkarni}, {Mooley}, {Kong}, {Cantwell}, {Yu}, {Cao}, {Fremling},
  {Neill}, {Ngeow}, {Nugent}, \& {Wozniak}}]{gezari_iPTF_2017}
{Gezari}, S., {Hung}, T., {Cenko}, S.~B., {et~al.} 2017, \apj, 835, 144,
  \dodoi{10.3847/1538-4357/835/2/144}

\bibitem[{{Greenwell} {et~al.}(2021){Greenwell}, {Gandhi}, {Stern}, {Boorman},
  {Toba}, {Lansbury}, {Mainieri}, \& {Desira}}]{greenwell_candidate_2021}
{Greenwell}, C., {Gandhi}, P., {Stern}, D., {et~al.} 2021, \mnras, 503, L80,
  \dodoi{10.1093/mnrasl/slab019}

\bibitem[{Hao {et~al.}(2007)Hao, Weedman, Spoon, Marshall, Levenson, Elitzur,
  \& Houck}]{hao_distribution_2007}
Hao, L., Weedman, D.~W., Spoon, H. W.~W., {et~al.} 2007, The Astrophysical
  Journal, 655, L77, \dodoi{10.1086/511973}

\bibitem[{{Harrison} {et~al.}(2013){Harrison}, {Craig}, {Christensen},
  {Hailey}, {Zhang}, {Boggs}, {Stern}, {Cook}, {Forster}, {Giommi},
  {Grefenstette}, {Kim}, {Kitaguchi}, {Koglin}, {Madsen}, {Mao}, {Miyasaka},
  {Mori}, {Perri}, {Pivovaroff}, {Puccetti}, {Rana}, {Westergaard}, {Willis},
  {Zoglauer}, {An}, {Bachetti}, {Barri{\`e}re}, {Bellm}, {Bhalerao},
  {Brejnholt}, {Fuerst}, {Liebe}, {Markwardt}, {Nynka}, {Vogel}, {Walton},
  {Wik}, {Alexander}, {Cominsky}, {Hornschemeier}, {Hornstrup}, {Kaspi},
  {Madejski}, {Matt}, {Molendi}, {Smith}, {Tomsick}, {Ajello}, {Ballantyne},
  {Balokovi{\'c}}, {Barret}, {Bauer}, {Blandford}, {Brandt}, {Brenneman},
  {Chiang}, {Chakrabarty}, {Chenevez}, {Comastri}, {Dufour}, {Elvis}, {Fabian},
  {Farrah}, {Fryer}, {Gotthelf}, {Grindlay}, {Helfand}, {Krivonos}, {Meier},
  {Miller}, {Natalucci}, {Ogle}, {Ofek}, {Ptak}, {Reynolds}, {Rigby},
  {Tagliaferri}, {Thorsett}, {Treister}, \& {Urry}}]{harrison_nuclear_2013}
{Harrison}, F.~A., {Craig}, W.~W., {Christensen}, F.~E., {et~al.} 2013, \apj,
  770, 103, \dodoi{10.1088/0004-637X/770/2/103}

\bibitem[{{HI4PI Collaboration} {et~al.}(2016){HI4PI Collaboration}, {Ben
  Bekhti}, {Fl{\"o}er}, {Keller}, {Kerp}, {Lenz}, {Winkel}, {Bailin},
  {Calabretta}, {Dedes}, {Ford}, {Gibson}, {Haud}, {Janowiecki}, {Kalberla},
  {Lockman}, {McClure-Griffiths}, {Murphy}, {Nakanishi}, {Pisano}, \&
  {Staveley-Smith}}]{HI4PI_collab_2016}
{HI4PI Collaboration}, {Ben Bekhti}, N., {Fl{\"o}er}, L., {et~al.} 2016, \aap,
  594, A116, \dodoi{10.1051/0004-6361/201629178}

\bibitem[{{Hickox} \& {Alexander}(2018)}]{hickox_obscured_2018}
{Hickox}, R.~C., \& {Alexander}, D.~M. 2018, \araa, 56, 625,
  \dodoi{10.1146/annurev-astro-081817-051803}

\bibitem[{{Hviding} {et~al.}(2018){Hviding}, {Hickox}, {Hainline}, {Carroll},
  {DiPompeo}, {Yan}, \& {Jones}}]{hviding_characterizing_2018}
{Hviding}, R.~E., {Hickox}, R.~C., {Hainline}, K.~N., {et~al.} 2018, \mnras,
  474, 1955, \dodoi{10.1093/mnras/stx2849}

\bibitem[{{Kollatschny} {et~al.}(2020){Kollatschny}, {Grupe}, {Parker},
  {Ochmann}, {Schartel}, {Herwig}, {Komossa}, {Romero-Colmenero}, \&
  {Santos-Lleo}}]{kollatschny_optical_2020}
{Kollatschny}, W., {Grupe}, D., {Parker}, M.~L., {et~al.} 2020, \aap, 638, A91,
  \dodoi{10.1051/0004-6361/202037897}

\bibitem[{{Kormendy} \& {Ho}(2013)}]{kormendy_coevolution_2013}
{Kormendy}, J., \& {Ho}, L.~C. 2013, \araa, 51, 511,
  \dodoi{10.1146/annurev-astro-082708-101811}

\bibitem[{Lamastra {et~al.}(2009)Lamastra, Bianchi, Matt, Perola, Barcons, \&
  Carrera}]{lamastra_bolometric_2009}
Lamastra, A., Bianchi, S., Matt, G., {et~al.} 2009, \aap, 504, 73,
  \dodoi{10.1051/0004-6361/200912023}

\bibitem[{{Madsen} {et~al.}(2015){Madsen}, {Harrison}, {Markwardt}, {An},
  {Grefenstette}, {Bachetti}, {Miyasaka}, {Kitaguchi}, {Bhalerao}, {Boggs},
  {Christensen}, {Craig}, {Forster}, {Fuerst}, {Hailey}, {Perri}, {Puccetti},
  {Rana}, {Stern}, {Walton}, {J{\o}rgen Westergaard}, \&
  {Zhang}}]{madsen_calibration_2015}
{Madsen}, K.~K., {Harrison}, F.~A., {Markwardt}, C.~B., {et~al.} 2015, \apjs,
  220, 8, \dodoi{10.1088/0067-0049/220/1/8}

\bibitem[{Maiolino {et~al.}(2001)Maiolino, Marconi, Salvati, Risaliti,
  Severgnini, Oliva, La~Franca, \& Vanzi}]{maiolino_dust_2001}
Maiolino, R., Marconi, A., Salvati, M., {et~al.} 2001, \aap, 365, 28,
  \dodoi{10.1051/0004-6361:20000177}

\bibitem[{Moran {et~al.}(2002)Moran, Filippenko, \&
  Chornock}]{moran_hidden_2002}
Moran, E.~C., Filippenko, A.~V., \& Chornock, R. 2002, \apj, 579, L71,
  \dodoi{10.1086/345314}

\bibitem[{{Murphy} \& {Yaqoob}(2009)}]{murphy_x-ray_2009}
{Murphy}, K.~D., \& {Yaqoob}, T. 2009, \mnras, 397, 1549,
  \dodoi{10.1111/j.1365-2966.2009.15025.x}

\bibitem[{{Nandra}(2006)}]{nandra_on_2006}
{Nandra}, K. 2006, \mnras, 368, L62, \dodoi{10.1111/j.1745-3933.2006.00158.x}

\bibitem[{Padovani {et~al.}(2017)Padovani, Alexander, Assef, De~Marco, Giommi,
  Hickox, Richards, Smolčić, Hatziminaoglou, Mainieri, \&
  Salvato}]{padovani_active_2017}
Padovani, P., Alexander, D.~M., Assef, R.~J., {et~al.} 2017, \aapr, 25, 2,
  \dodoi{10.1007/s00159-017-0102-9}

\bibitem[{Reyes {et~al.}(2008)Reyes, Zakamska, Strauss, Green, Krolik, Shen,
  Richards, Anderson, \& Schneider}]{reyes_space_2008}
Reyes, R., Zakamska, N.~L., Strauss, M.~A., {et~al.} 2008, \aj, 136, 2373,
  \dodoi{10.1088/0004-6256/136/6/2373}

\bibitem[{{Ricci} {et~al.}(2016){Ricci}, {Bauer}, {Arevalo}, {Boggs}, {Brandt},
  {Christensen}, {Craig}, {Gandhi}, {Hailey}, {Harrison}, {Koss}, {Markwardt},
  {Stern}, {Treister}, \& {Zhang}}]{ricci_IC751_2016}
{Ricci}, C., {Bauer}, F.~E., {Arevalo}, P., {et~al.} 2016, \apj, 820, 5,
  \dodoi{10.3847/0004-637X/820/1/5}

\bibitem[{{Ricci} {et~al.}(2017{\natexlab{a}}){Ricci}, {Trakhtenbrot}, {Koss},
  {Ueda}, {Del Vecchio}, {Treister}, {Schawinski}, {Paltani}, {Oh}, {Lamperti},
  {Berney}, {Gandhi}, {Ichikawa}, {Bauer}, {Ho}, {Asmus}, {Beckmann}, {Soldi},
  {Balokovi{\'c}}, {Gehrels}, \& {Markwardt}}]{ricci_bat_2017}
{Ricci}, C., {Trakhtenbrot}, B., {Koss}, M.~J., {et~al.} 2017{\natexlab{a}},
  \apjs, 233, 17, \dodoi{10.3847/1538-4365/aa96ad}

\bibitem[{{Ricci} {et~al.}(2017{\natexlab{b}}){Ricci}, {Trakhtenbrot}, {Koss},
  {Ueda}, {Schawinski}, {Oh}, {Lamperti}, {Mushotzky}, {Treister}, {Ho},
  {Weigel}, {Bauer}, {Paltani}, {Fabian}, {Xie}, \&
  {Gehrels}}]{ricci_close_2017}
---. 2017{\natexlab{b}}, \nat, 549, 488, \dodoi{10.1038/nature23906}

\bibitem[{{Risaliti} {et~al.}(2005){Risaliti}, {Elvis}, {Fabbiano}, {Baldi}, \&
  {Zezas}}]{risaliti_rapid_2005}
{Risaliti}, G., {Elvis}, M., {Fabbiano}, G., {Baldi}, A., \& {Zezas}, A. 2005,
  \apjl, 623, L93, \dodoi{10.1086/430252}

\bibitem[{{Schawinski} {et~al.}(2015){Schawinski}, {Koss}, {Berney}, \&
  {Sartori}}]{schawinski_active_2015}
{Schawinski}, K., {Koss}, M., {Berney}, S., \& {Sartori}, L.~F. 2015, \mnras,
  451, 2517, \dodoi{10.1093/mnras/stv1136}

\bibitem[{{Shen} \& {Ho}(2014)}]{shen_diversity_2014}
{Shen}, Y., \& {Ho}, L.~C. 2014, \nat, 513, 210, \dodoi{10.1038/nature13712}

\bibitem[{{Shu} {et~al.}(2010){Shu}, {Yaqoob}, \& {Wang}}]{shu_cores_2010}
{Shu}, X.~W., {Yaqoob}, T., \& {Wang}, J.~X. 2010, \apjs, 187, 581,
  \dodoi{10.1088/0067-0049/187/2/581}

\bibitem[{Stern(2015)}]{stern_x-ray_2015}
Stern, D. 2015, \apj, 807, 129, \dodoi{10.1088/0004-637X/807/2/129}

\bibitem[{{Stern} {et~al.}(2018){Stern}, {McKernan}, {Graham}, {Ford}, {Ross},
  {Meisner}, {Assef}, {Balokovi{\'c}}, {Brightman}, {Dey}, {Drake},
  {Djorgovski}, {Eisenhardt}, \& {Jun}}]{stern_mid-ir_2018}
{Stern}, D., {McKernan}, B., {Graham}, M.~J., {et~al.} 2018, \apj, 864, 27,
  \dodoi{10.3847/1538-4357/aac726}

\bibitem[{{Vignali} {et~al.}(2006){Vignali}, {Alexander}, \&
  {Comastri}}]{vignali_quest_2006}
{Vignali}, C., {Alexander}, D.~M., \& {Comastri}, A. 2006, \mnras, 373, 321,
  \dodoi{10.1111/j.1365-2966.2006.11033.x}

\bibitem[{{Vignali} {et~al.}(2010){Vignali}, {Alexander}, {Gilli}, \&
  {Pozzi}}]{vignali_disovery_2010}
{Vignali}, C., {Alexander}, D.~M., {Gilli}, R., \& {Pozzi}, F. 2010, \mnras,
  404, 48, \dodoi{10.1111/j.1365-2966.2010.16275.x}

\bibitem[{{Wachter} {et~al.}(1979){Wachter}, {Leach}, \&
  {Kellogg}}]{wachter_parameter_1979}
{Wachter}, K., {Leach}, R., \& {Kellogg}, E. 1979, \apj, 230, 274,
  \dodoi{10.1086/157084}

\bibitem[{Wright {et~al.}(2010)Wright, Eisenhardt, Mainzer, Ressler, Cutri,
  Jarrett, Kirkpatrick, Padgett, McMillan, Skrutskie, Stanford, Cohen, Walker,
  Mather, Leisawitz, Gautier, McLean, Benford, Lonsdale, Blain, Mendez, Irace,
  Duval, Liu, Royer, Heinrichsen, Howard, Shannon, Kendall, Walsh, Larsen,
  Cardon, Schick, Schwalm, Abid, Fabinsky, Naes, \&
  Tsai}]{wright_wide-field_2010}
Wright, E.~L., Eisenhardt, P. R.~M., Mainzer, A.~K., {et~al.} 2010, The
  Astronomical Journal, 140, 1868, \dodoi{10.1088/0004-6256/140/6/1868}

\bibitem[{{Wu} {et~al.}(2012){Wu}, {Brandt}, {Anderson}, {Diamond-Stanic},
  {Hall}, {Plotkin}, {Schneider}, \& {Shemmer}}]{wu_x-ray_2012}
{Wu}, J., {Brandt}, W.~N., {Anderson}, S.~F., {et~al.} 2012, \apj, 747, 10,
  \dodoi{10.1088/0004-637X/747/1/10}

\bibitem[{{Yuan} {et~al.}(2016){Yuan}, {Strauss}, \&
  {Zakamska}}]{yuan_spectroscopic_2016}
{Yuan}, S., {Strauss}, M.~A., \& {Zakamska}, N.~L. 2016, \mnras, 462, 1603,
  \dodoi{10.1093/mnras/stw1747}

\end{thebibliography}
\bibliographystyle{aasjournal}

\end{document}